\documentclass[aps,floatfix,superscriptaddress]{revtex4}
\pdfoutput=1 
\usepackage{amsmath,amssymb,eucal,graphicx,epsfig,epstopdf}

\newcommand{\eq}[1]{(\ref{#1})}
\newcommand{\fig}[1]{Fig.~\ref{#1}}

\newcommand{\be}{\begin{equation}}
\newcommand{\ee}{\end{equation}}

\newcommand{\ve}{\mathbf}

\begin{document}
\title{Geometrical selection in growing needles}

\author{P. L. Krapivsky}
\affiliation{Department of Physics, Boston University, Boston, Massachusetts 02215, USA}

\author{L. I. Nazarov}
\affiliation{Physics Department, Moscow State University, 119992, Moscow, Russia}

\author{ M. V. Tamm}
\affiliation{Physics Department, Moscow State University, 119992, Moscow, Russia}
\affiliation{Department of Applied Mathematics, MIEM, National Research University Higher School of Economics, 123458, Moscow, Russia}


\begin{abstract}
We investigate the growth of needles from a flat substrate. We focus on the situation when needles suddenly begin to grow from the seeds randomly distributed on the line. The width of needles is ignored and we additionally assume that (i) the growth rate is the same for all needles; (ii) the direction of the growth of each needle is randomly chosen from the same distribution; (iii) whenever the tip of a needle hits the body of another needle, the former needle freezes, while the latter continues to grow. We elucidate the large time behavior by employing an exact analysis and the Boltzmann equation approach. We also analyze the evolution when seeds are located on a half-line, on a finite interval. Needles growing from the two-dimensional substrate are also examined. 
\end{abstract}

\maketitle

\section{Introduction}

Crystallization of a substance from vapor or melt often implies simultaneous formation of many crystallite seeds on a substrate. These seeds than continue to grow from the substrate competing for the material left in the vapor or melt. Such crystals are subject to geometrical selection---the probability for a given seed to grow into a large crystal depends on its geometrical orientation \cite{lamm,kolm,vdDrift}. In the case of growth of thin needle crystals (see \cite{apatite,GaN,obraz1,obraz2,obraz3} for examples of different substances growing in needles), geometrical selection essentially implies that when two thin needle crystals meet, the less `vertical' needle stops growing while the more vertical needle continues to grow.

In this paper we propose a toy model of geometrical selection and analyze its properties in various settings including growth from 1D or 2D substrate with uniform density of initial seeds, growth from a finite set of seeds, and growth from seeds distributed with constant density on a half-line. In the case of infinite 1D substrate we show the problem is closely related to the one-dimensional traffic model \cite{traffic} allowing us to find an exact solution in the totally asymmetric case and to construct exact upper and lower bounds on the probability of a needle survival in the general case. We also solve the Boltzmann equation and show that it gives qualitatively correct decay laws, e.g. the asymptotic density of surviving needles is correct up to an amplitude, while more subtle features like the decay of the needles substantially more tangential to the substrate than the typical surviving needles are wrong. In the 2D case, we analyze the scaling behavior and the Boltzmann equation. When the number of needles growing from the 1D substrate is finite, we show that the exact distribution of the number of ultimately surviving needles is the same as the distribution of the number of surviving clusters in the 1D ballistic aggregation \cite{IBA1,IBA2,IBA3,majumdar}. In the case of the initially occupied half-line, we calculate the large time asymptotic of the average number of needles infiltrating the seed-free half-line. 


The paper is organized as follows. We define the model in section \ref{sec:model}. In section \ref{sec:VD}, we study the angle distribution for growing needles on an infinite 1D substrate. In section \ref{sec:2D}, we construct a scaling theory for the angle distribution for growing needles on the 2D substrate and we analyze the corresponding Boltzmann equation. In section \ref{sec:in}, we present several exact results and conjectures concerning the needles growing from a finite part of the 1D substrate. In section \ref{sec:conc}, we summarize our results and discuss a few unsolved problems. 

\section{The Model}
\label{sec:model}

We mimic a needle by a ray growing from a seed. Seeds are randomly distributed on the one-dimensional horizontal line $y=0$, and the needles grow into the upper half-plane $y>0$. The speed of the growth is assumed to be the same for all needles (we set it equal to unity). The direction of growth is random, and this causes the interaction between the needles---one must define what happens when the tip of one needle hits the body of another one. We postulate that such a collision freezes the first needle, while the second needle is not affected. 

In the realm of our model the evolution is fully deterministic, the randomness is only in the initial conditions, and the interaction between the needles occurs only in collisions. A number of different needle growth models have been investigated, see \cite{Krug-rev} for a review. In these models the growth mechanism was usually stochastic, and the interactions were also very different (e.g., caused by some kind of screening or shadowing mechanism). For instance, Laplacian needles where the interaction is via a Laplacian field have been studied in Refs.~\cite{Lap-Meakin,Lap-Rossi,Lap-Rossi-2,Lap-Hakim,Lap-Krug}. Needle models in which the growth is caused by ballistic deposition were also studied, particularly by Krug and Meakin \cite{Krug-Meakin,Krug-Meakin-1,Krug-Meakin-2}.

\begin{figure}[t]
\includegraphics[width=17cm]{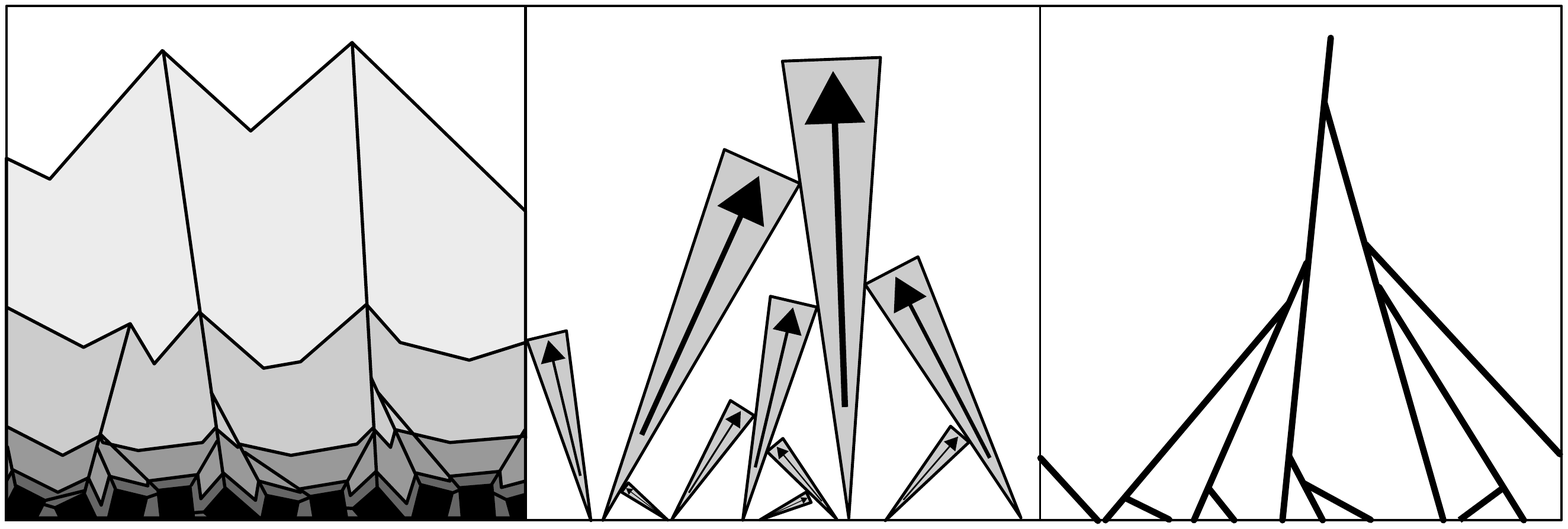}
\vspace{-2in}
\caption{(a) Schematic illustration of the formation of a polycrystalline film by geometric selection, in case of crystals growing from seeds of arbitrary polygon shape (adopted from \cite{vdDrift}); (b) similar geometric selection for thin pyramidal crystals (adopted from \cite{obraz2}); (c) the limiting case of arbitrary thin needles considered in this paper. In all cases the selection rule on collisions is that the more vertical line survives while the less vertical perishes.}
\label{collision_fig}
\end{figure}

The analysis of this strongly interacting infinite-needle system simplifies after projection on the one-dimensional horizontal line from which the growth has begun. We then follow the motion of the projections of the tips. Each such projection, a particle, moves with a certain velocity $v$; in terms of the inclination angle $\theta$ of its direction of growth to the vertical axis, $|\theta|\leq \tfrac{\pi}{2}$, the velocity is $v=\sin\theta$. We assume that (i) initial velocities are uncorrelated and drawn from the same velocity distribution $P_0(v)$; (ii) 
initial positions (the locations of the seeds) are also uncorrelated, without loss of generality we set the density to unity.

Many of our results are valid for an arbitrary $P_0(v)$. Symmetric velocity distributions, $P_0(v)=P_0(-v)$, usually arise in applications and some of the results simplify in this case. Therefore, we often consider symmetric velocity distributions. The results also significantly simplify for totally asymmetric velocity distributions, $P_0(v)=0$ for $v<0$, as we shall see below. 

Let $P(v,t)$ be the velocity distribution of growing needles. This quantity contains the total density 
\begin{equation}
\label{density_def}
n(t)=\int_{-1}^1 dv\,P(v,t)
\end{equation}
and the average velocity 
\begin{equation}
\label{speed_def}
\langle |v|\rangle = \frac{1}{n(t)} \int_{-1}^1 dv\,|v|P(v,t)\,.
\end{equation}
Needless to say, $P_0(v)\equiv P_0(v, t=0)$. Since we set the initial density to unity:
$n(t=0)=\int_{-1}^1 dv\,P_0(v)=1$.

At first sight, the description in terms of the point particles moving on the one-dimensional line looks simpler than the original description it terms of needles growing in the two-dimensional space. The collision rule becomes more complicated, however: Colliding particles have different positions. Indeed, in a collision of particles with velocities of the same sign, the particles never meet (the fast particle disappears before catching the slow particle); in a collision of particles moving toward each other, they pass through each other (the actual collision of needles occurs later). Overall in every collision, the particle moving with larger speed disappears (see Fig.~\ref{collision_fig}c). 

Consider two particles with initial coordinates $x_1$ and $x_2$ and velocities $v_1$ and $v_2$ and assume, for definiteness, that $x_1<x_2$ and $v_1>v_2>0$. The tip of the first needle will hit the second needle at a certain time $t$. The tip of the second needle was there at some earlier time $\tau$. We have 
\begin{equation*}
t \cos\theta_1 = \tau \cos\theta_2, \quad t \sin\theta_1 = \tau \sin\theta_2 + x_2 - x_1\,.
\end{equation*}
Recalling that $v_1 = \sin\theta_1$ and $v_2 = \sin\theta_2$, we find
\begin{equation}
\label{interval}
\tau= t\sqrt{\frac{1-v_1^2}{1-v_2^2}}\,, \quad 
x_2-x_1= t \!\left(v_1-v_2\sqrt{\frac{1-v_1^2}{1-v_2^2}}\right).
\end{equation}
The initial distance between the tips is $x_2-x_1$. The final distance between the tips (after the projection on the horizontal line) is 
\begin{equation}
\label{final}
x_2-x_1+t(v_2-v_1)=(x_2-x_1)v_2\,\frac{\sqrt{1-v_2^2}-\sqrt{1-v_1^2}}{v_1\sqrt{1-v_2^2}-v_2\sqrt{1-v_1^2}}\,.
\end{equation}
It is easy to check that the same equations hold when the velocities of the two colliding particles have different sign (i.e., $v_1>0>v_2$, and $|v_1|>|v_2|$.


\section{Velocity Distribution} 
\label{sec:VD}

To determine the velocity distribution $P(v,t)$ of growing needles we employ the method developed in the context of traffic model \cite{traffic}, see also a textbook exposition \cite{book}. In the traffic model the collision occurs when two particles (representing cars) are at the same place; in the needle model the tips are at different places, see \eqref{final}. Further, in the traffic problem the proper initial velocity distribution $P_0(v)$ was totally asymmetric, $P_0(v)=0$ for $v<0$. This was crucial in deriving the exact velocity distribution \cite{traffic}
\begin{equation}
\label{pvt_traffic}
P(v,t) = P_0(v)\, \exp\!\left[-t\int_{0}^v dw\,P_0(w)\!\left(v-w\right)\right].
\end{equation}
The needle problem with totally asymmetric velocity distribution ($P_0(v)=0$ for $v<0$) is also exactly solvable. The corresponding velocity distribution
\begin{equation}
\label{pvt_asym}
P(v,t) = P_0(v) \exp\!\left[-t\int_{0}^v dw\,P_0(w)\!\left(v-w\sqrt{\frac{1-v^2}{1-w^2}}\right)\right]
\end{equation}
is the direct generalization of \eq{pvt_traffic} for the non-local rules of needle collision, see the paragraph below \eq{pvt} for the derivation of this result. We have not succeeded in finding $P(v,t)$ for an arbitrary initial velocity distribution, and even for symmetric initial velocity distributions. In the general case it is possible to establish upper and lower bounds on the distribution $P(v,t)$ as we now demonstrate. 

\subsection{Exact Bounds} 

Consider a target particle moving with velocity $v$ ($v>0$ for concreteness). This particle can be eliminated in a collision with a slower particle on the right of the target particle. In order for a slower particle with velocity $w, \, |w|<v$, to be able to eliminate the target particle before time $t$, the slow particle should be located within a distance 
\begin{equation}
\label{delta_x}
\Delta x (v,w; t) = t\!\left(v-w\sqrt{\frac{1-v^2}{1-w^2}}\right)
\end{equation}
of the target particle. If for any $|w|<v$ there is no potential `killer' particle in the corresponding interval $\Delta x(v,w; t)$, the target particle is guaranteed to survive up to time $t$. This gives the following lower bound $P_{\ell}$ \cite{plus}
\begin{equation}
\label{pvt}
P(v,t)\! \geq \! P_{\ell}(v,t) \!= \! P_0(v) \exp\!\left[-t\int_{-v}^v dw\,P_0(w)\!\left(v-w\sqrt{\frac{1-v^2}{1-w^2}}\right)\right].
\end{equation}
The exponential reflects the assumption that the initial seeds are randomly located (with unit density); the length of the interval around the target particle which must be free of slow particles is given by \eqref{delta_x}. 

\begin{figure}[t]
\vspace{-0.4in}
\includegraphics[width=17cm]{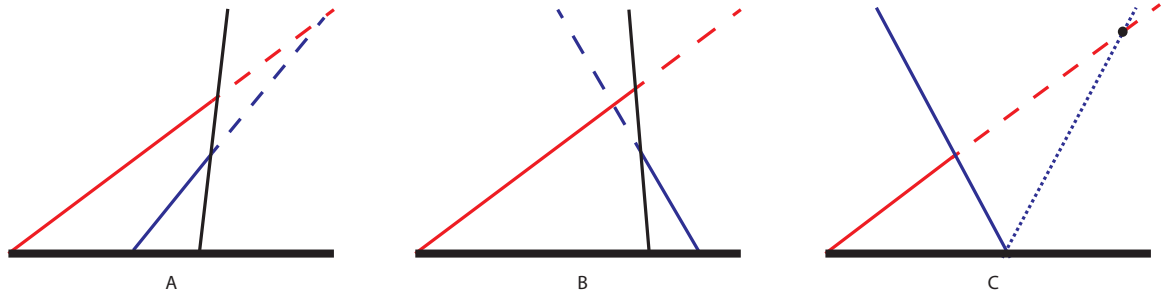}
\vspace{-2.2in}
\caption{Possible interplay of three intersecting needles. (a) In the case of needles with positive inclinations the existence of the third (black) needles leads to the red needle being killed even earlier then it would have been if only the blue one existed; (b) this is not true if needles can have both positive and negative inclinations: in this case the black needle might kill the red one later than the blue one would have done; (c) still, one can write an upper bound on the needle surviving time: the existence of the blue needle guarantees that the red one will be dead beyond the black point (see explanation in the text).}
\label{Fig:New}
\end{figure}

When the velocities of {\it all} particles are positive, the lower bound \eqref{pvt} is simultaneously an upper bound, and hence the exact solution. Indeed, it is possible that the killer particle (blue in Fig.~\ref{Fig:New}a) does not eliminate the target particle (red in Fig.~\ref{Fig:New}a) because there exists another particle (`killer'-2, black in the figure) which eliminates it before it reaches the red one. But if all particle velocities are positive, the black particle is guaranteed to eliminate the red one even earlier than the blue one. As a result, in this case the existence of the blue particle within the interval $\Delta x(v,w; t)$ guarantees that the red one will be dead either at the moment $t$ or earlier.

This logic is inapplicable when velocities can be both positive and negative: \fig{Fig:New}b shows an example of a particle which is eliminated {\it later} than it would have been by the initial (blue) killer particle. However, one can still construct an upper bound for the survival probability. Consider the situation with the target (red) and the killer (blue) particles having velocities with different signs (\fig{Fig:New}c). Because of possible interference of other particles, we cannot guarantee that the red particle will be dead after its intersection with the world line of the blue particle. However, the target particle cannot survive beyond the black point in \fig{Fig:New}c, which is the intersection of the world line of the target particle and of a virtual particle which starts in the same point as the blue one, and has the velocity $-w$, i.e., opposite to the velocity of the blue particle. Indeed, it is obvious that any possible third particle which could possibly eliminate the blue one would intersect with the world line of red particle before the black point. The same is true for the possible fourth particle which could eliminate the third, and so on {\it ad infinitum}. Thus, we arrive at the upper bound
\begin{equation}
\label{pvt2}
P(v,t)\! \leq \! P_u(v,t) \! = \! P_0(v)\exp\!\left[-t\int_{-v}^v dw\,P_0(w)\!\left(v-|w|\sqrt{\frac{1-v^2}{1-w^2}}\right)\right]
\end{equation}

The bounds \eqref{pvt}--\eqref{pvt2} are valid for an arbitrary velocity distribution. For a completely asymmetric velocity distribution, $P_0(v)=0$ for $v<0$, the two bounds coincide giving the announced exact solution \eqref{pvt_asym}. 

In the remaining part of this subsection we consider only symmetric velocity distributions, $P_0(v)=P_0(-v)$. For such distributions, the bounds \eqref{pvt}--\eqref{pvt2} give 
\begin{equation}
\label{pvt_sym}
e^{-2 t \left[I_1(v)- I_2(v)\right]} \geq \frac{P(v,t)}{P_0(v)} \geq e^{-2 t I_1(v)}
\end{equation}
with
\begin{subequations}
\begin{align}
\label{I_1}
I_1 (v) &= v \int_0^v dw\,P_0(w) \\
\label{I_2}
I_2 (v)& = \int_0^v dw\, w\sqrt{\frac{1-v^2}{1-w^2}}\,P_0(w)
\end{align}
\end{subequations}

For the uniform initial velocity distribution
\begin{equation}
\label{uniform}
P_0(v)=
\begin{cases}
\tfrac{1}{2} & |v|<1\\
0 & |v|>1
\end{cases}
\end{equation}
which corresponds to the situation when the initial distribution of the inclination angles is given by $\Pi(\theta)=\tfrac{1}{2}\cos\theta$ for $|\theta|\leq \tfrac{\pi}{2}$, the integrals \eq{I_1} and \eq{I_2} are $I_1 = v^2/2$ and $I_2 = \sqrt{1-v^2}\left(1-\sqrt{1-v^2}\right)$, so the bounds become
\begin{equation}
\label{pvt_uni}
\frac{1}{2}\exp \left[- t \left(1- \sqrt{1-v^2}\right)\right] \geq \! P(v,t) \geq \! \frac{1}{2}\exp\left[- v^2t\right]
\end{equation}

For the uniform initial distribution of the inclination angles, $\Pi(\theta)=\pi^{-1}$ for $|\theta|\leq \tfrac{\pi}{2}$, the initial velocity distribution is $P_0(v)=\pi^{-1}(1-v^2)^{-1/2}$. In this case $I_1 = \pi^{-1} v\, \sin^{-1}v$ and $I_2 = - (2\pi)^{-1} \ln (1-v^2)$ leading to 
\begin{equation}
\label{pvt_2}
\exp\!\left[-\frac{2tv\sin^{-1}v}{\pi}\right] (1-v^2)^{-t/\pi}\geq \! \frac{P(v,t)}{P_0(v)} \geq \! \exp\!\left[-\frac{2tv\sin^{-1}v}{\pi}\right]
\end{equation}

The long-time behavior of the velocity distribution is determined by the small-velocity behavior of the initial velocity distribution. Assuming an algebraic small velocity behavior,
\begin{equation}
P_0(v)\simeq A |v|^\mu 
\label{mu}
\end{equation}
when $|v|\ll 1$, one gets 
\begin{equation}
I_1\simeq \frac{A}{\mu+1} |v|^{\mu+2} ,\quad I_2\simeq \frac{A}{\mu+2} |v|^{\mu+2}
\end{equation}
allowing one to write the exact bounds \eqref{pvt}--\eqref{pvt2} in the scaling form
\begin{equation}
\label{Pvt_scaling}
A|v|^\mu\, \exp\!\left[-B_1|v|^{\mu+2}t\right] \geq \! P(v,t) \geq \! A|v|^\mu\, \exp\!\left[-B_2|v|^{\mu+2}t\right]\,, 
\end{equation}
where
\be
B_1=\frac{2A}{(\mu+1)(\mu+2)}\,, \qquad B_2 = \frac{2A}{\mu+1}
\ee
Using \eqref{Pvt_scaling} we find that the density and the average speed exhibit simple algebraic behaviors in the large time limit:
\begin{subequations}
\begin{align}
\label{n-mu}
& 2 A(1-\nu)\Gamma(\nu) \left(B_2 t\right) ^{-\nu}\, \geq \,n \,\geq\, 2 A(1-\nu)\Gamma(\nu) \left(B_1 t\right) ^{-\nu}\\
\label{v-mu}
&\frac{1}{\Gamma(1-\nu)}\,\left(B_2 t\right)^{\nu-1}\geq \langle |v|\rangle \geq \frac{1}{ \Gamma(1-\nu)}\,\left(B_1 t\right)^{\nu-1}
\end{align}
\end{subequations}
where we used short-hand notation 
\be
\nu=\frac{\mu+1}{\mu+2}
\label{nu}
\ee

The velocity distribution is expected to acquire a scaling form
\begin{equation}
\label{Pvt-scal}
P(v,t)=A t^{-\mu/(\mu+2)} V^\mu \mathcal{P}(V), \quad V = |v|\, t^{1/(\mu+2)}
\end{equation}
in the long-time limit. The bounds \eqref{Pvt_scaling} show that $\mathcal{P}(0)=1$ and imply that $\ln \mathcal{P}(V) \simeq -BV^{\mu+2}$ as $V\gg 1$. The exact value of $B$ is unknown; Eqs.~\eqref{Pvt_scaling} lead to the bounds $B_1\leq B\leq B_2$.

The most natural case from the point of view of the needle growth problem is one where initial velocity distribution remains finite at zero velocity, i.e., $\mu=0$. In this case the velocity distribution becomes
\begin{equation}
\label{Pvt_0}
A \exp\!\left[-Av^2t\right]\geq P(v,t) \geq A \exp\!\left[-2Av^2t\right]
\end{equation}
while the particle density and the average particle speed both decay as $t^{-1/2}$:
\begin{equation}
\label{nv_exact}
\sqrt{\frac{\pi A}{t}} \geq n \geq \sqrt{\frac{\pi A}{2t}}\,, \quad \frac{1}{\sqrt{\pi A t}} \geq \langle |v|\rangle \geq \frac{1}{\sqrt{2\pi A t}}.
\end{equation}
These asymptotic behaviors agree with predictions of Kolmogorov \cite{kolm} who arrived to them using scaling arguments; Kolmogorov also gave hints about exact amplitudes which he planned to derive in a later (never published) work. The $t^{-1/2}$ decay of the density of needles has been also found in a number of needle models in which the underlying growth mechanism is stochastic, see e.g. \cite{Krug-Meakin,Krug-Meakin-1}.

\subsection{Boltzmann Equation Approach} 


In this subsection we consider only symmetric velocity distributions, $P_0(-v)=P_0(v)$. In this situation, the lower bound \eqref{pvt} simplifies to 
\begin{subequations}
\begin{equation}
\label{pvt:L}
P_{\ell}(v,t) = P_0(v) \exp\!\left[-vt\int_{-v}^v dw\,P_0(w)\right]. 
\end{equation}
Equivalently, the lower bound can be obtained by solving a linear Boltzmann-like equation
\begin{equation}
\label{BE_exact}
\frac{\partial P_{\ell}(v,t)}{\partial t}=-vP_{\ell}(v,t)\int_{-v}^v dw\,P_0(w).
\end{equation}
\end{subequations}
In turn, the upper bound \eqref{pvt2} which we re-write as
\begin{subequations}
\begin{equation}
\label{pvt2:U}
P_u(v,t) = P_0(v)\exp\!\left[-vt\int_{-v}^v dw\,P_0(w)\!\left(1-\sqrt{\frac{v^{-2}-1}{w^{-2}-1}}\right)\right]
\end{equation}
satisfies a linear Boltzmann-like equation
\begin{equation}
\label{BE_exact_U}
\frac{\partial P_{u}(v,t)}{\partial t} = -v P_{u}(v,t) \int_{-v}^v dw\,P_0(w)\!\left(1-\sqrt{\frac{v^{-2}-1}{w^{-2}-1}}\right)
\end{equation}
\end{subequations}
The Boltzmann equation describing the evolution of $P(v,t)$ has the form
\begin{equation}
\label{BE-naive}
\frac{\partial P(v,t)}{\partial t}=-P(v,t)\int_{-v}^v dw\, \left(v-w\sqrt{\frac{1-v^2}{1-w^2}}\right) P(w,t)=-vP(v,t)\int_{-v}^v dw\,P(w,t),
\end{equation}
where we have used \eq{interval} and simplified the integral by using the symmetry: $P(-w,t)=P(w,t)$. In contrast to the linear Boltzmann-like equations \eqref{BE_exact} and \eqref{BE_exact_U} for the lower and upper bounds, the Boltzmann equation \eqref{BE-naive} is a non-linear integro-differential equation. One can reduce \eqref{BE-naive} to a non-linear partial differential equation (PDE)
\begin{equation}
\label{BE-naive-PDE}
v\,\frac{\partial^2 P}{\partial v \partial t} - \frac{v}{P}\,\frac{\partial P}{\partial v}\,\frac{\partial P}{\partial t} 
= \frac{\partial P}{\partial t} - 2 v^2 P^2
\end{equation}
which looks even more challenging that \eqref{BE-naive}. 
The Boltzmann equation \eqref{BE-naive} and its PDE version \eqref{BE-naive-PDE} appear analytically intractable. In the most interesting long-time limit, however, one can derive asymptotically exact results since the solution acquires a scaling form. We now describe the procedure in the most natural case of a continuous symmetric velocity distribution which remains finite at zero velocity, $P(v=0)=A>0$. The structure of Eq.~\eqref{BE-naive} suggests to seek the velocity distribution in the scaled form
\begin{equation}
\label{scaled}
P(v,t) = A F(V), \quad V=v\, \sqrt{2At}
\end{equation}
Plugging \eqref{scaled} into Eq.~\eqref{BE-naive} we obtain
\begin{equation}
\label{F}
F' = - 2F(V)\int_0^V dW\,F(W)
\end{equation}
where $(\cdot)'=d(\cdot)/dV$. Writing $\Phi(V) = \int_0^V dW\,F(W)$ one transforms \eqref{F} into $\Phi'' = -2\Phi\Phi'$ which can be further integrated to yield $\Phi' = 1 - \Phi^2$ and then $\Phi = \tanh(V)$ leading to 
\begin{equation}
\label{P_Boltzmann} 
F_\text{BE}(V) = \frac{1}{\cosh^2(V)}
\end{equation}
Clearly, this scaled velocity distribution does not lie within the exact bounds $e^{-V^2/2} \geq F_\text{exact}(V) \geq e^{-V^2}$ [see \eqref{Pvt_0}]. 
Equations \eqref{scaled} and \eqref{P_Boltzmann} predict the following asymptotic decay of the density and the average speed
\begin{equation}
\label{nv_BE}
n_\text{BE} = \sqrt{\frac{2 A}{t}}\,, \quad \langle |v|\rangle_\text{BE} = \frac{\ln 2}{\sqrt{2A t}}.
\end{equation}
Comparing with the exact bounds \eqref{nv_exact} we conclude that the Boltzmann equation approach gives qualitatively correct $t^{-1/2}$ decay laws. The amplitudes predicted by the Boltzmann equation approach are probably erroneous, although they lie inside the exact bounds \eqref{nv_exact}. 
\section{Two Dimensions}
\label{sec:2D}

In this section we consider needles of finite width growing from a two-dimensional substrate. We start with needles of constant width and then consider a more realistic model of conical needles with linearly growing width.

\subsection{Cylindrical Needles (Constant Width)}

Here we assume that the needles are cylinders with equal radii $a$. Other assumptions are like in the one-dimensional setting, e.g. the speed of the longitudinal growth is assumed to be the same for all needles and the growth begins at the same time $t=0$. In the long time limit the projections of the surviving needles will have small velocities. The molecular chaos assumption underlying the Boltzmann equation is plausible in two dimensions, while in one dimension it is violated as we have seen in the previous section. 

\begin{figure}[h]
\includegraphics[width=0.55\textwidth,clip=]{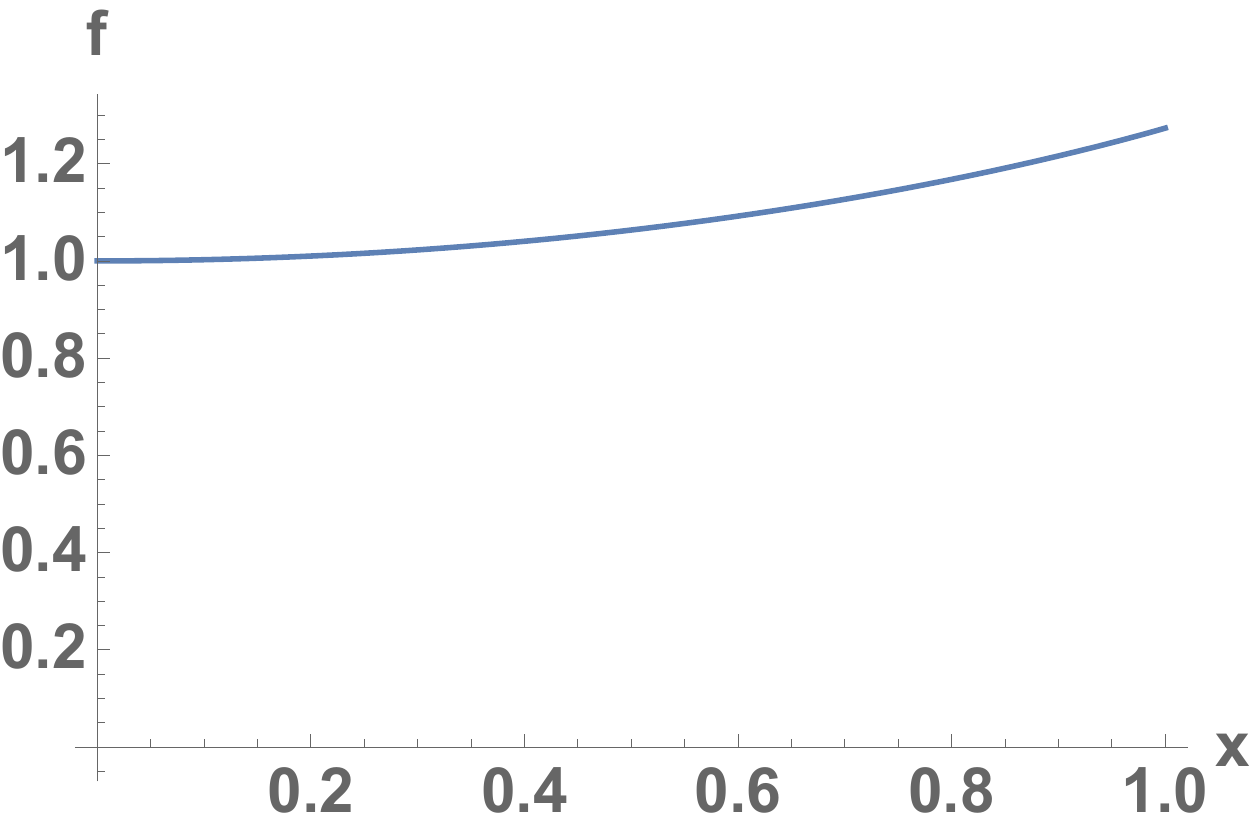}
\caption{The function $f(x)$ appearing in \eqref{BE_2d} is function of $x$ monotonically increasing from $f(0)=1$ 
to $f(1)=\frac{4}{\pi}$.}
\label{Fig:fx}
\end{figure}

Within the Boltzmann equation approach, the velocity distribution evolves according to 
\begin{equation}
\label{BE_2d_general}
\frac{\partial P(\ve v,t)}{\partial t}=-a \int_{|\ve w|<|\ve v|} P(\ve v,t) P(\ve w,t) \left|\ve v - \ve w\sqrt{\frac{1-v^2}{1-w^2}}\right| d\ve w. 
\end{equation}
We limit ourselves by the symmetric case when the velocity distribution depends only on the speed $v = |\ve v|$. After changing to the polar coordinates, one gets
\be
\frac{\partial P(\ve v,t)}{\partial t}=-a \int_{0}^v P(v,t) P(w,t) w dw\, \int_0^{2\pi} d\phi \sqrt{X(v,w) -Y(v,w) \cos \phi},
\label{BE_2d_2}
\ee
where $\phi$ is the angle between vectors $\ve v$ and $\ve w$, and 
\be
X(v,w) = \frac{v^2+w^2 - 2 w^2 v^2}{1-w^2};\;\; Y(v,w) = 2 v w \sqrt{\frac{1-v^2}{1-w^2}}.
\label{XY}
\ee
In the case of small velocities ($v \ll 1$) which is the only one relevant at large time, one can drop higher order terms in \eq{XY} and rewrite \eq{BE_2d_2} in a more convenient form
\begin{equation}
\label{BE_2d}
\frac{\partial P(v,t)}{\partial t}=- 2\pi a v^3 P (v,t) \int_0^1 dx\,x f(x) P(xv,t)
\end{equation}
where we have used the shorthand notation 
\begin{equation}
\label{fx:def}
f(x) = \frac{1}{2\pi} \int_0^{2\pi} d\phi \,\sqrt{1+x^2-2x \cos \phi}\,. 
\end{equation}
The function $f(x)$ can be expressed through the complete elliptic integral of the second kind. From such an exact expression, or directly from \eqref{fx:def}, one finds that $f(x)$ slowly increases as $x$ increases, more precisely (see Fig.~\ref{Fig:fx})
\begin{equation*}
f(x) = 
\begin{cases}
1+\frac{x^2}{4}+\frac{x^4}{64}+\ldots & 0<x\ll 1\\
\frac{4}{\pi}-\frac{2(1-x)}{\pi}+\ldots & 0<1-x\ll 1
\end{cases}
\end{equation*}

We again assume that $P(v=0)=A>0$ and seek the velocity distribution in the scaling form \eqref{scaled}. The scaling of the velocity should be modified, however. The structure of Eq.~\eqref{BE_2d} shows that $t^{-1}\sim Aav^3$, so the properly scaled velocity distribution is
\begin{equation}
\label{scaled_2d}
P(v,t) = A F(V), \quad V=v \sqrt[3]{2\pi Aat} 
\end{equation}
Making this substitution we transform the Boltzmann equation \eqref{BE_2d} into
\begin{equation}
\label{F_2d}
F' = - 3 F V^2 \int_0^1 dx\,x f(x) F(xV) 
\end{equation}
which should be solved subject to the boundary condition
\begin{equation}
\label{BC}
F(0) = 1
\end{equation}

The boundary-value problem \eqref{F_2d}--\eqref{BC} is analytically intractable, so we limit ourselves to asymptotic behaviors. In the small velocity limit, $0<V\ll 1$, the scaled velocity distribution admits an expansion
\begin{equation}
\label{FV:small}
F(V) = 1- \alpha V^3 + \beta V^6+\ldots 
\end{equation}
(Perhaps a bit surprisingly, the scaling function $F(V)$ is non-analytical in the $V=|\ve V| \to 0$ limit.) By inserting the expansion \eqref{FV:small} into \eqref{F_2d} we extract the amplitudes
\begin{subequations}
\begin{align}
\label{alpha}
&\alpha = \int_0^1 dx\,x f(x) = \frac{16}{9\pi} \\
\label{beta}
&\beta = \frac{\alpha}{2}\left[\alpha+\int_0^1 dx\,x^4 f(x)\right] = \frac{128}{81\pi^2}+ \frac{845+18G}{1296\pi^2}
\end{align}
\end{subequations}
where $G = \sum_{k\geq 0}(-1)^k/(2k+1)^{-2}$ is the Catalan constant. In the large velocity limit, $V\gg 1$, one easily establishes an exponential decay
\begin{equation}
F(V) \propto e^{-CV}, \quad C= 3\int_0^\infty dW\,WF(W). 
\end{equation}
It does not seem possible to find an analytical expression for $C$. The scaling form \eqref{scaled_2d} gives the decay laws for the density and the average speed
\begin{equation}
\label{nv_2d}
n \sim t^{-2/3}\,, \quad \langle |v|\rangle \sim t^{-1/3}
\end{equation}

\subsection{Conical Needles (Growing Width)}

We now assume that needles are (circular) cones with radii growing linearly with time (distance from the seed), as shown in Fig. 1b. There is some similarity with the touch-and-stop model of growth \cite{ABK1,ABK2,Dodds1,Dodds2}, also known as the lilypond model \cite{lily1,lily2,lily3}, although in that model every collision freezes both disks.

The Boltzmann equation reads
\begin{equation}
\label{BE_cone}
\frac{\partial P(v,t)}{\partial t}=-\lambda t v^3 P(v,t)\int_0^1 dx\,x P(vx,t) f(x),
\end{equation}
which is nothing but \eq{BE_2d} with time-dependent $a$. The scaled velocity distribution has the form
\begin{equation}
\label{scaled_cone}
P(v,t) = A F(V), \quad V=(\lambda A t^2)^{1/3} v
\end{equation}
The decay laws for the density and the average speed are 
\begin{equation}
\label{nv_cone}
n \sim t^{-4/3}\,, \quad \langle |v|\rangle \sim t^{-2/3}
\end{equation}
The scaled velocity distribution satisfies the same equation \eqref{F_2d} (apart from a different numerical factor in the right-hand side), and therefore the large-velocity tail is again exponential
\begin{equation}
\label{tail}
F(V) \sim \exp\left[ - CV\right]
\end{equation}

One can similarly analyze other growth laws of the radii of cones. Indeed, if the radius of a cone grows as $a \sim t^{\alpha}$ we must replace $\lambda t\rightarrow \lambda t^{\alpha}$ on the RHS of \eqref{BE_cone}. The proper scaling variable is now 
\be
V\sim v t^{(1+\alpha)/3}.
\ee 
Hence the decay laws for the particle density and the average particle speed are
\be
n \sim t^{-2(1+\alpha)/3};\;\; \langle |v|\rangle \sim t^{-(1+\alpha)/3}.
\label{nv_general}
\ee

\begin{figure}
\centerline{
\includegraphics[width=5cm]{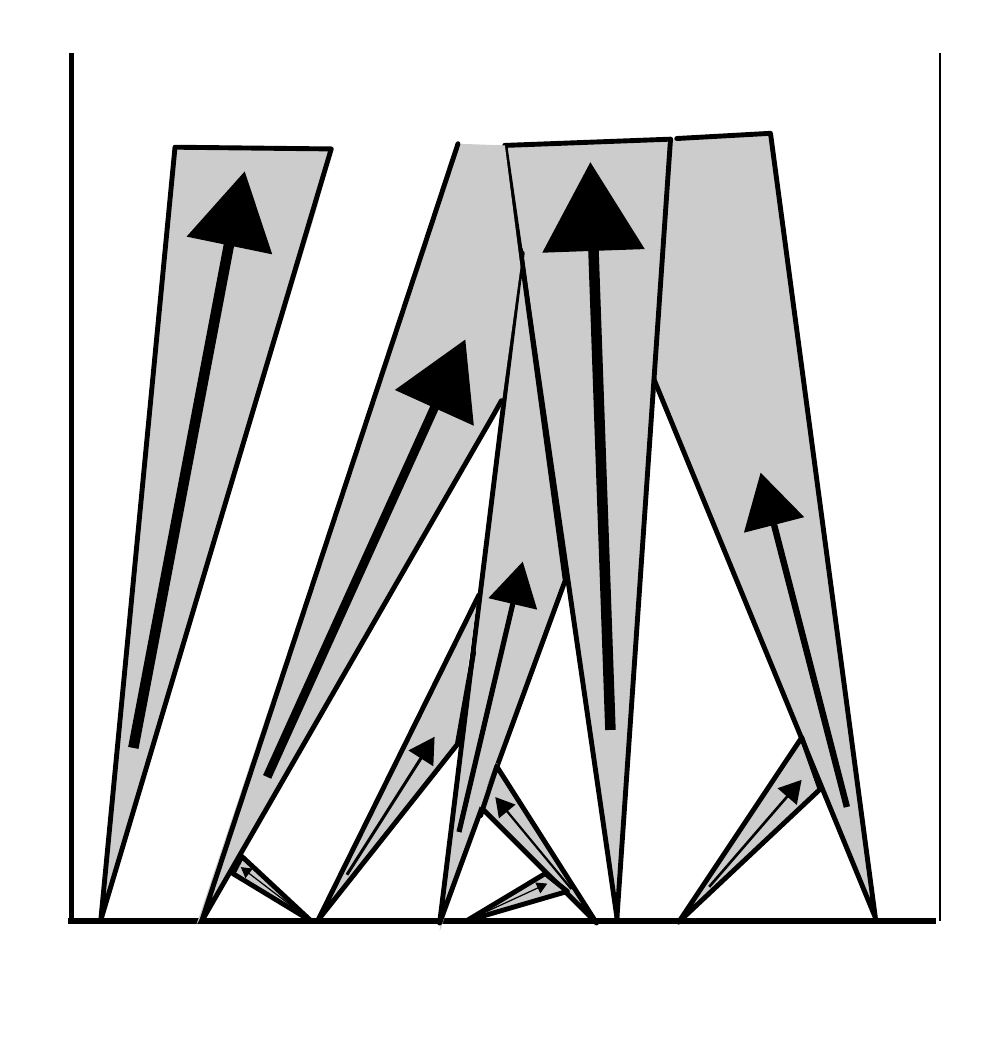}}
\caption{Inosculating needles, i.e. needles which are not dying immediately after collision.}
\label{fig_film}
\end{figure}

The consideration above implies that a needle (either cylindrical or conical) stops growing immediately after collision with a more vertical one. In a more realistic scenario, such needles do not perish immediately---the side of the needle in touch with a more vertical one stops growing, while the other side `does not know' that a collision happened and continues growing (see \fig{fig_film}). As a result, a continuous film of inosculated needles is formed. Continuity implies that the typical transversal size of the surviving part of the needle scales as $n^{-1/2}$. This in conjunction with \eq{nv_general} yields
\be
a \sim t^{\alpha} \sim \left( t^{-2(1+\alpha)/3}\right)^{-1/2} \; \rightarrow \; \alpha = 1/2. 
\ee 
Thus $n \sim t^{-1}$ and $\langle |v|\rangle \sim t^{-1/2}$ for the case of inosculating needles. These scaling laws were stated by Kolmogorov \cite{kolm} who apparently relied on similar scaling arguments. 

\section{Needles growing from an interval or half-line} 
\label{sec:in}

In this section we return to the one-dimensional substrate and consider what happens if the original substrate is not uniformly covered by seeds. First, we consider the evolution of a finite number of needles and calculate the distribution of the number of immortal needles, i.e. needles surviving up to $t = \infty$. Second, we discuss the set-up when needle seeds are located on a half-line and study how they infiltrate the half-line which is initially free of needles.

\subsection{Finite number of needles}

Consider the evolution of $N$ needle seeds initially located at a certain finite interval. Our aim is to calculate the probability distribution $\Pi(n|N)$ of exactly $n$ of them surviving ad infinitum. We assume that needle velocities are taken independently from the same distribution $P_0(v)$, with symmetric $P_0(v)$ being the most interesting case.

\begin{figure}
\includegraphics[width=14cm]{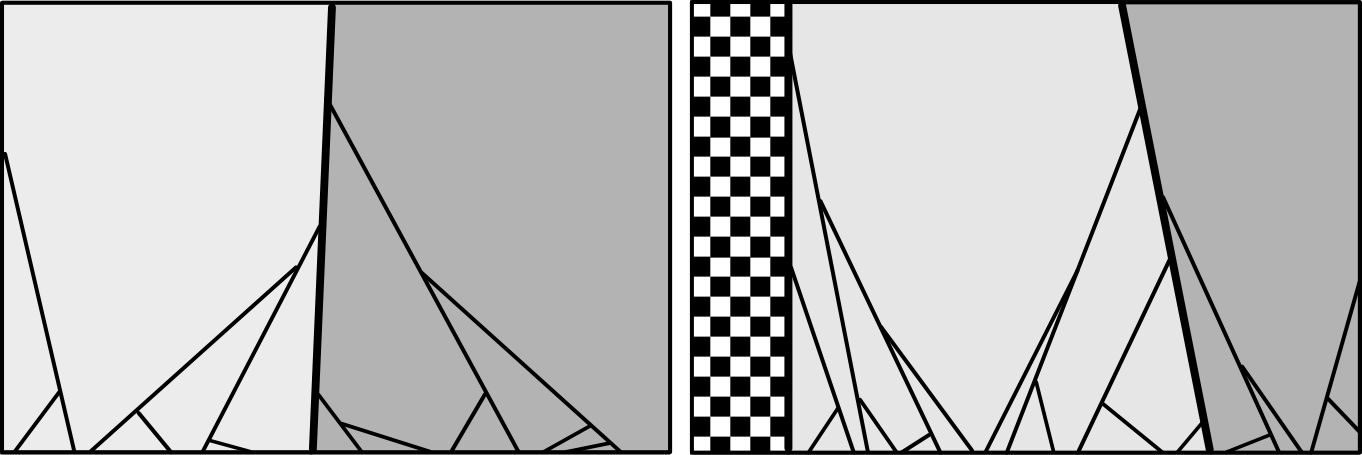}
\caption{Recurrence relations for the survival probabilities in the system with finite number of needles. (A): The most vertical needle (thick line) separates the whole set of needles into two domains (shaded differently) which evolve separately; (B) needles in half-line: the most vertical needle separates the set into part confined between itself in the wall (lightly shaded) where all the needles eventually perish, and the outer part (densely shaded) for which it works as an effective new wall. }
\label{fig:finite_rec}
\end{figure}

The set of $N$ needles can be recursively separated into smaller subsets which have almost no interactions with one another and this allows us to construct recursive relations for $\Pi(n|N)$ as we show below. To appreciate this idea, consider the needle with the smallest speed (we call it `slow needle-1' below). Note that, (i) this needle is immortal---there is no other slower needle to kill it, (ii) the needles on the left and on the right of it will never interact, and (iii) the survival of other needles does not depend on the velocity of this slowest needle: any other needle moving towards it will eventually perish. This allows us to write 
\begin{equation}
\Pi(n|N) = \sum_{M=0}^{N-1} \sum_{m=0}^{n-1} \frac{1}{N} \Pi_R(m|M) \Pi_L(n-m-1|N-M-1) ,
\label{fan_rec}
\end{equation}
We label needles from right to left and denote by $M$ the number of needles to the right of the slowest; the $1/N$ factor is the probability that $(M+1)$-th needle is the slowest. $\Pi_{R,L}(m|M)$ are the probabilities that exactly $m$ out of $M$ needles to the right (respectively, left) of the slowest needle are ultimate survivors (see \fig{fig:finite_rec} for an illustration). If $P_0(v)$ is symmetric, $\Pi_R(m|M)=\Pi_L(m|M)$. 

The distributions $\Pi_{R,L}(m|M)$ satisfy similar recurrences. Take for instance $\Pi_R(m|M)$. The slowest needle of the set of $M$ needles (`slow needle-2') separates it into two non-interacting subsets. The subset to the left of it will definitely die out: indeed, it is contained between slow needle-1 and slow needle-2, which work effectively as absorbing walls, while the subset to the right forms a new rightmost subset for which slow needle-2 plays exactly the same role as slow needle-1 for the original set. Finally, the survival of the slow needle-2 itself depends only on the sign of its velocity: if it is positive it will survive ad infinitum, if it is negative it will collide with slow needle-1 and die. Collecting all this, and taking into account that every needle of the set has equal probability to be the slowest, we get
\be
\Pi_R(m|M) = \frac{1}{M} \sum_{K=0}^{M-1} [p_L \Pi_R(m|K) + p_R\Pi_R(m-1|K)].
\label{one-side_rec}
\ee
Here $p_L$ (respectively, $p_R$) is the probability that the slowest needle has negative (respectively, positive) velocity. The recurrence relation for $\Pi_{L}$ is the same, up to replacement of indexes: $R \to L$. Needless to say, $p_L = p_R = 1/2$ for symmetric $P_0(v)$, and $p_L = 0, p_R = 1$ for the completely asymmetric one. In a more general case when $P_0(-v)/P_0(v)$ is a $v$-independent constant $\lambda$, we have $p_L = \lambda/(\lambda+1)$ and $p_R = 1/(\lambda+1)$. For a generic $P_0(v)$, the probabilities $p_L$ and $p_R$ depend on the velocity of the needle in question.

Initial and boundary conditions read 
\be
\Pi_R(m|0)=\delta_{m,0}, \quad \Pi_R(-1|M)=0 
\ee
We introduce generating functions 
\begin{subequations}
\begin{align}
\label{G-gen}
& G(s,z) = \sum_{M=0}^{\infty} \sum_{m=0}^{\infty} \Pi(m|M) s^m z^M \\
\label{GRL-gen}
&G_{R,L}(s,z) = \sum_{M=0}^{\infty} \sum_{m=0}^{\infty} \Pi_{R,L}(m|M) s^m z^M
\end{align}
\end{subequations}
The one-sided recursion \eq{one-side_rec} can be re-written as
\be
\sum_{M=0}^{\infty} \sum_{m=0}^{\infty} M \Pi_R(m|M) s^m z^M= (p_L + s p_R) \sum_{m=0}^{\infty} \sum_{K=0}^{\infty} \Pi_R(m|K) \sum_{M=K+1}^{\infty} s^m z^M,
\label{one-side_gen1}
\ee
which reduces to
\be
z\frac{\partial G_R(s,z)}{\partial z} = (p_L + s p_R) \frac{z}{1-z} G_R(s,z)
\label{one-side_gen2}
\ee
Integrating this equation, one gets for a one-sided generating function
\be
G_R(s,z) = (1-z)^{- (p_L + s p_R)}\,.
\label{one-side_gen3}
\ee
Similarly, $G_L(s,z)=(1-z)^{- (p_R + s p_L)}$. 

The two-sided recursion equation \eq{fan_rec} turns into equation 
\be
z \frac{\partial G(s,z)}{\partial z} = s z G_R(s,z) G_L(s,z)
\label{fan_gen1}
\ee
for the generating functions. Thus 
\be
G(s,z) = s \int \frac{dz}{(1-z)^{p_L + s p_R} (1-z)^{p_R + s p_L}}=s \int \frac{dz}{(1-z)^{1 + s}}=(1-z)^{-s},
\label{fan_gen2}
\ee
which is a well-known \cite{knuth,flajolet} generating function of the Stirling numbers \cite{Stirling} of the first kind:
\be
\sum_{N\geq 0}\sum_{n=0}^N 
\begin{bmatrix}
N\\
n
\end{bmatrix}
s^n\, \frac{z^N}{N!} = (1-z)^{-s}
\label{stirling}
\ee
Therefore,
\be
\Pi(n|N) = \frac{1}{N!}
\begin{bmatrix}
N\\
n
\end{bmatrix}
\label{fan_result}
\ee

It is now easy calculate the moments of the $\Pi(n|N)$ distribution exploiting the properties of the Stirling numbers. 
For example, the average and the variance of the number of ultimately surviving needles 
\be
\left< n \right> = H_N, \quad \left< n^2 \right> -\langle n \rangle^2 = H_N - H_N^{(2)}
\ee
are expressed through harmonic numbers $H_N=\sum_{1\leq k\leq N}k^{-1}$ and $H_N^{(2)}=\sum_{1\leq k\leq N}k^{-2}$. 

The same distribution \eq{fan_result} describes the outcome of the ballistic aggregation process where particles undergo totally inelastic collisions \cite{IBA1,IBA2,IBA3,majumdar}. Another example is a bullet problem \cite{annihilation} where $N$ bullets are shot one after another in the same direction and whenever two bullets collide they both annihilate---in that problem the emerging distribution $\Pi_{ann}(n|N)$ is rather similar to \eq{fan_result}. The distribution \eq{fan_result} also often arises in theory of records \cite{Arnold,Nevzorov}, in studies of lead changes in networks \cite{KR02,JML1}, and in numerous problems in combinatorics \cite{knuth,flajolet,theater}. 
\subsection {Infiltration}

Consider now the situation when the seeds are uniformly distributed on the half-line $x<0$, namely, the probability density $P(x,v, t=0)$ to find a seed at $x$ with velocity $v$ is
\begin{equation}
P(x,v, t=0) = 
\begin{cases}
P_0(v) & x \leq 0\\
0 & x>0.
\end{cases}
\end{equation}

In this case, similarly to the case of a finite set of seeds, some needles survive ad infinitum. For the $n^\text{th}$ needle from the boundary, we denote by $s_n$ its ultimate survival probability. The $n^\text{th}$ needle is eternal if it has a positive velocity which is smaller than the speeds of all $n-1$ needles to the right of it. If $P_0(v)$ has a fixed asymmetry, i.e. 
\be
\frac{P_0(-v)}{P_0(v)} = \lambda
\ee
for all $v>0$, the sign and the speed are independent random variables and
\be
s_n = \frac{\Lambda}{n}\,, \quad \Lambda \equiv \frac{1}{1+\lambda} 
\label{sn}
\ee
where $\Lambda$ is the probability that velocity is positive, and $n^{-1}$ is a probability that absolute value of velocity is the smallest among $n$ equally distributed ones. In particular, $s_n = 1/(2n)$ for symmetric distributions, and $s_n = 1/n$ for completely asymmetric ones. For more general $P_0(v)$, the ultimate survival probability $s_n$ converges to \eq{sn} when $n\gg 1$ if we define 
\be 
\lambda = \lim_{v \to 0} \frac{P_0(-v)}{P_0(v)} \,. 
\label{limlambda}
\ee
Moreover, given \eq{scaled} one expects that probability of survival at a finite time $\tilde{s}_n(t)$ behaves as
\be
\tilde{s}_n (t) = s_n + a_n t^{-\nu} + o (t^{-\nu}),
\ee
where $\nu$ is defined by Eq.~\eq{nu}.

We also want to determine the probability $r_n$ that the $n^\text{th}$ needle is not only immortal but also becomes the rightmost at infinite time, i.e., that it survives, but all needles on the right of it are eliminated. In the completely asymmetric case the rightmost needle always survives, $r_n = \delta_{1,n}$; in the general case, the rightmost needle always survives 
only if its velocity is positive. It is instructive to introduce a cumulative distribution
\be
R_0 =1, \quad R_n = 1 - \sum_{k=1}^{n} r_k \quad \text{ for }\quad n\geq 1,
\ee
which is the probability that all $n$ rightmost needles eventually perish. The $n^\text{th}$ needle will be the rightmost at $t = \infty$, if (i) it survives till infinite time; (ii) needles to the right of it are eliminated. For the distributions with fixed asymmetry these two events are independent---the event (ii) depends only on the sign of the velocities of the first $(n-1)$ needles, which are completely decoupled from absolute values of the velocities in this case. Thus one gets
\begin{equation}
\label{R-rec}
r_n = R_{n-1} - R_{n} = s_n R_{n-1}, \quad R_n = (1-s_n) R_{n-1}.
\end{equation}
Recalling \eqref{sn} we solve the above recurrence and obtain 
\begin{equation}
R_n = \frac{\Gamma(n+1-\Lambda)}{\Gamma(1-\Lambda)\,\Gamma(n+1)}\,, \quad 
r_n = \frac{\Lambda}{\Gamma(1-\Lambda)}\, \frac{\Gamma(n-\Lambda)}{\Gamma(n+1)} 
\end{equation}
which show that $R_n \sim n^{-\Lambda}$ and $r_n \sim n^{-1-\Lambda}$ when $n\gg 1$.

\begin{figure}
\includegraphics[width=14cm]{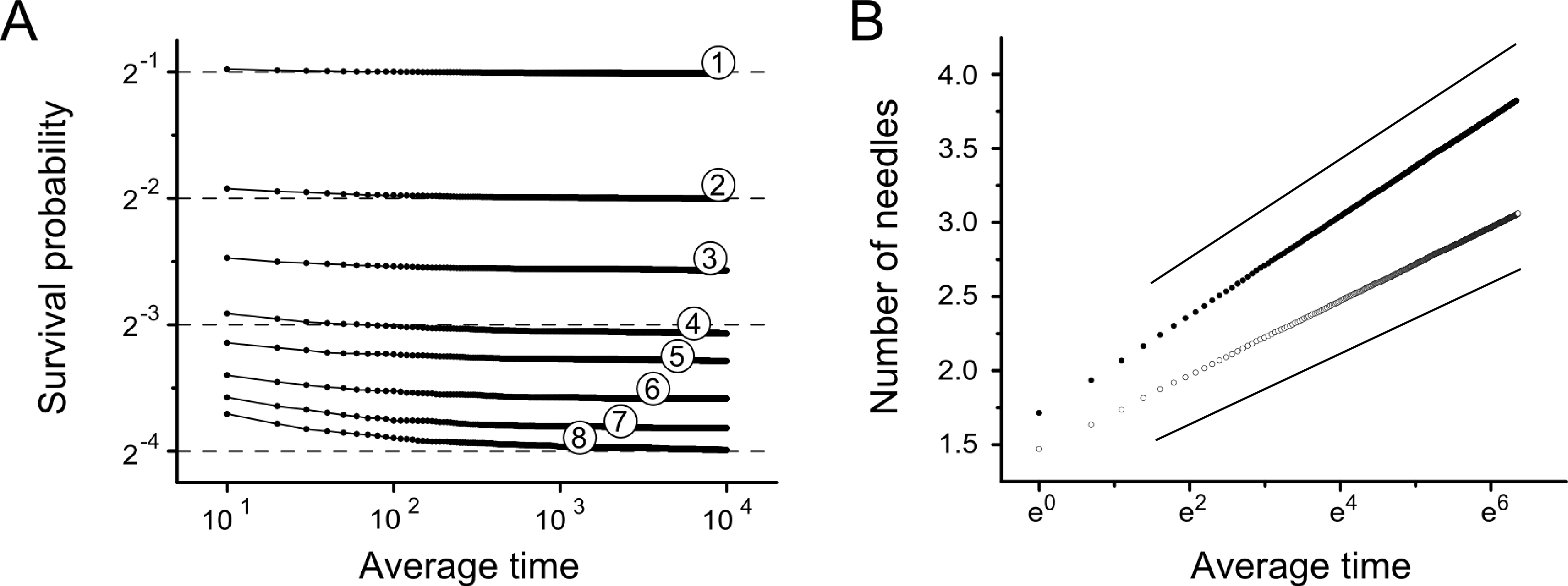}
\caption{Numerical simulations of the needle infiltration. (A) The survival probability of $n^\text{th}$ needle as a function of time for $n=1...8$ converges to $1/2n$ as $t \to \infty$. (B) The average number $N(t)$ of alive needles infiltrating the positive half-line at time $t$ (open circles), and the total number $M(t)$ of needles which have infiltrated the positive half-line up to time $t$ including those which eventually froze (disks). The straight lines are the guides for the eyes and have slopes $1/4$ and $1/3$.}
\label{fig:infiltration}
\end{figure}

Let us look at needles that infiltrate initially empty half-line $x>0$. One interesting quantity is the total number of such needles which are alive at time $t$:
\be
N(t) = \int_0^{\infty}dx \int_0^{\infty}dv P(x,v, t).
\ee 
Another interesting quantity is the total number of needles which infiltrated the positive half-line {\it up to} time $t$:
\be
M(t)= \int_0^{t} d\tau \int_0^{\infty}dv \left.\frac{\partial P(x,v,\tau)}{\partial x}\right|_{x=0}
\ee 
In \fig{fig:infiltration} we show the numerical results for $N(t)$ and $M(t)$ in the case of symmetric uniform distribution \eq{uniform}. Both quantities grow logarithmically with slopes suspiciously close to 1/4 and 1/3, respectively.

We have not determined $N(t)$ or $M(t)$ analytically, but we have computed a similar quantity $I(t)$, the average number of `immortal' needles, which infiltrate the positive half-line at time $t$. Clearly, $I(t) \leq N(t) \leq M(t)$. To determine $I(t)$ we first compute the survival probability $S(v,x)$ of a needle starting at point $-x$ with velocity $v$:
\be
S(v,x) = \Theta(v) \exp \left( - x \int_{-v}^{v} P_0(w) dw \right),
\ee
where $\Theta (v)$ is the Heaviside function. Then $I(t)$ is simply
\be
I(t) = \int_0^{\infty} dv \int_0^{vt} dx P_0(v) S(v,x) = \int_0^{\infty} dv\,\frac{P_0(v)}{F(v)} \left[ 1- \exp\left(-v F(v) t\right) \right],
\label{immortal}
\ee
where $F(v) = \int_{-v}^{v} dw\, P_0(w)$. The long-time behavior of \eq{immortal} is controlled by the behavior of $P_0(v)$ at small velocities:
\be
P_0(v) \simeq \Lambda \frac{d F}{dv}\,, \quad F(v) \simeq a v^{\mu+1},
\ee
where $\mu$ and $\Lambda$ are defined by \eq{mu} and \eq{sn}, and $a$ is a numerical constant. This allows to rewrite \eq{immortal} in the form
\be
I(t) \simeq \Lambda \int_0^{Y(t)} \frac{1-\exp(-y)}{y}dy = \Lambda \frac{\mu+1}{\mu+2} \ln t + O(1),
\label{immortal2}
\ee
where $Y(t) = a^{-1/(\mu+2)} t^{(\mu+1)/(\mu+2)}$. In the most interesting case of symmetric velocity distribution with $P_0(0)>0$
\be
I(t) = \frac{1}{4} \ln t + O(1),
\label{immortal3}
\ee
Simulation results suggest (see \fig{fig:infiltration}B) that $N(t) - I(t) = O(1)$, i.e. the number of alive but mortal particles in the positive half-line remains finite throughout the evolution. The total number of particles $M(t)$ that penetrated into the positive half-line also grows logarithmically, but with a different pre-factor; when $\mu=0, \lambda=1$ the pre-factor looks suspiciously close to 1/3. Analytical calculation of the amplitude in the general case is an interesting challenge.

\section{Discussion}
\label{sec:conc}

For needles growing from the one-dimensional substrate, we have found upper and lower bounds for the velocity distribution. In the case of completely asymmetrical distributions these bounds coincide, making the problem exactly solvable. We have also solved the Boltzmann equation in the situation when the initial velocity distribution is symmetric and finite at zero velocity. The chief scaling laws for the average speed and the density of the surviving needles predicted by the Boltzmann equation approach are correct. The details are erroneous, e.g., the prediction for the velocity distribution is incompatible with exact bounds. This is not surprising---the Boltzmann equation approach is an uncontrolled approximation and is known to work poorly in the one-dimensional settings. 

We have also discussed the version of the problem when needles were seeded only on part of the line. In the case when the number of needles is finite, the distribution of the number of ultimately surviving needles is remarkably universal. Specifically, it coincides with the distribution of surviving particles in the ballistic aggregation process where particles undergo totally inelastic collisions \cite{IBA1,IBA2,IBA3,majumdar}. The universality of this answer seems to imply there should be some universal derivation not relying on the details of a particular model. A slightly different distribution arises in the context of the ballistic annihilation problem, although the derivation in this situation is much more involved \cite{annihilation}. It seems plausible that there exist wide classes of the one-dimensional aggregation and annihilation processes in which the distribution of the fan size is given by $\Pi(n|N)$ and $\Pi_{ann}(n|N)$, respectively. Determining the minimal collision rule requirements which lead to these distributions is an interesting problem to address. 

In two dimensions, we have employed the Boltzmann equation approach. The Boltzmann equation approach is more sound in two dimensions than in one dimension, yet it is still an uncontrolled approximation. For instance, whenever the tip of the faster cone hits the slower one, their radii are different. The temporal scaling is still the same, so the applicability of \eqref{BE_cone} seems plausible. It would be interesting to study the problem numerically and to check the validity of the theoretical prediction \eqref{tail} for the tail. 

If the lateral size of needles grows linearly in time, the needle model closely resembles the touch-and-stop, or the lilypond, growth model \cite{ABK1,ABK2,Dodds1,Dodds2,lily1,lily2,lily3}. In the case of the strictly vertical growth, whenever two objects touch their tips are at the the same height and hence both objects freeze, the needle model is isomorphic to the 1D lilypond which was exactly solved in \cite{ABK2,lily3}; in the case of the conical needles growing from the 2D substrate the model is isomorphic, again in the case of the strictly vertical growth, to the 2D lilypond model. The growing objects in \cite{ABK1,ABK2,Dodds1,Dodds2,lily1,lily2,lily3} are hyper-balls, but the lilypond model admits various modifications. One such version, a line-segment lilypond model \cite{lily4}, has needles of zero widths and assumes that the seeds are distributed in the plane $\mathbb{R}^2$, while in our model the seeds are on the 1D substrate. For a finite number of seeds studied in \cite{lily4} the distribution of the number of ultimately surviving needles is unknown. It would be also interesting to study the model \cite{lily4} with infinitely many seeds uniformly distributed throughout the plane and to determine the decay law for the fraction of mobile needles. A version of the line-segment lilypond model in which seeds are nucleated uniformly in space and time and grow only vertically or horizontally was proposed in Ref.~\cite{Rao} as a simple model describing martensites formation. The asymptotic behavior of this model is analytically tractable \cite{BK96}, but the analysis substantially uses the anisotropy of the growth.

\section*{Acknowledgements}

We thank A. N. Obraztsov and A. M. Alekseev for introducing us to the field of needle growth and L. N. Rashkovich for useful comments on the history of the subject. Large part of this work was done during several visits of MT and LN to the Applied Mathematics Research Center (AMRC), Coventry University. MT and LN are very grateful to the AMRC for the warm hospitality and to the project EU-FP7-PEOPLE-IRSES DIONICOS for financial support.


\begin{thebibliography}{99}
\bibitem{plus}
In Eq.~\eqref{pvt} and many following equations we tacitly assume that $v\geq 0$; the generalization
to $v<0$ is straightforward, and since we are mostly interested in symmetric distributions, it suffices
to know the behavior in the $v\geq 0$ region.
\end{thebibliography}
\end{document}